\documentclass[11pt]{article}
 \usepackage[letterpaper,hmargin=1in,vmargin=1.00in]{geometry}

\usepackage{hyperref}
\usepackage{latexsym,amssymb,amsmath,amsfonts,amscd,color,epsfig,amsthm}
\usepackage{pst-all}
\usepackage{tweaklist}
\renewcommand{\enumhook}{\setlength{\topsep}{0pt}  \setlength{\itemsep}{-3pt}}

\usepackage{float}

\floatstyle{ruled}
\newfloat{protocol}{tph}{lop}
\floatname{protocol}{Protocol} 
\providecommand{\abs}[1]{\lvert#1\rvert}

\newtheorem{theorem}{Theorem}[section]

\newtheorem{lemma}[theorem]{Lemma}

\begin{document}

\title{Information-theoretic security \\ without an honest
majority\\} \vspace{15pt}
\author{Anne Broadbent  and   Alain Tapp\\[.2cm]
\normalsize\sl D\'epartement d'informatique et de recherche op\'erationnelle\\[-0.1cm]
\normalsize\sl Universit\'e de Montr\'eal, C.P.~6128, Succ.\ Centre-Ville\\[-0.1cm]
\normalsize\sl Montr\'eal (QC), H3C 3J7~~\textsc{Canada}\\[.2cm]
\url{{broadbea, tappa}@iro.umontreal.ca} }

\date{June 13, 2007}

\maketitle

\begin{abstract}We present six multiparty protocols with
information-theoretic security that tolerate an arbitrary number of
corrupt participants. All protocols assume pairwise authentic
private channels and a broadcast channel (in a single case, we
require a simultaneous broadcast channel). We give protocols for
\emph{veto},  \emph{vote}, \emph{anonymous bit transmission},
\emph{collision detection}, \emph{notification} and \emph{anonymous
message transmission}. Not assuming an honest majority, in most
cases, a single corrupt participant can make the protocol abort. All
protocols achieve functionality never obtained before without the
use of either computational assumptions or of an honest majority.
\end{abstract}

\section{Introduction}

In the most general case, \emph{multiparty secure computation}
enables~$n$ participants to collaborate to compute
an~$n$-input,~$n$-output function (one per participant). Each
participant only learns his private output which, depending on the
function, can be the same for each participant. Assuming that
private random keys are shared between each pair of participants, we
known that every  function can be securely computed in the presence
of an active adversary if and only if less than~$n/3$ participants
are corrupt; this fundamental result is due to Michael Ben-Or, Shafi
Goldwasser and Avi Wigderson~\cite{BGW88} and David Chaum, Claude
Cr\'epeau and Ivan Damg\a rard~\cite{CCD88}. When a  broadcast
channel is available, the results of Tal Rabin and Michael
\mbox{Ben-Or}~\cite{RB89} tell us that this proportion can be
improved to~$n/2$.

Here, we present six specific multiparty computation protocols that
achieve  correctness  and privacy \emph{without} any assumption on
the number of corrupt participants. Naturally, we cannot always
achieve the ideal functionality, for example in some cases, a single
participant can make the protocol abort. This is the price to pay to
tolerate an arbitrary number of corrupt participants and still
provide information-theoretic privacy of the inputs.

All protocols we propose have polynomial complexity in the number of
participants and the security parameter. We always assume pairwise
shared private random keys between each pair of participants, which
allows pairwise private authentic channels. We also assume a
broadcast channel and, even though it is a strong assumption,  in
some cases we need the broadcast to be
simultaneous~\cite{CGMW85,HD05}.

\subsection{Summary of results}

Our main contributions are in the areas of elections (\emph{vote})
and anonymity (\emph{anonymous bit transmission} and \emph{anonymous
message transmission}). Each protocol is an astute combination of
basic protocols, which are also of independent interest, and that
implement \emph{parity}, \emph{veto}, \emph{collision detection} and
\emph{notification}.

The main ingredient for our information-theoretically  secure
protocols is the dining cryptographers protocol~\cite{Chaum88} (see
also Section~\ref{sec:parity}), to which we add the following simple
yet powerful observation: if~$n$ participants each hold a private
bit of an~$n$-bit string with Hamming weight of parity~$p$, then any
single participant can randomize~$p$ by locally flipping his bit
with a certain probability. It is impossible, however, for any
participant to locally derandomize~$p$. In the case of the anonymous
message transmission, we also build on the dining cryptographers
protocol by noting that a message that is sent  can be ciphered with
a one-time pad by having one participant (the receiver) broadcast a
random bit. Any modification of the message can then be detected by
the receiver with an \emph{algebraic manipulation detection
code}~\cite{CFP07}.

\subsubsection{Vote}

Our \emph{vote} protocol (Section~\ref{sec:vote}) allows~$n$
participants to conduct an $m$-candidate election. The privacy is
perfect but the protocol has the drawback that if it aborts (any
corrupt participant can cause an abort), the participants can still
learn information that would have been available had the protocol
succeeded.  For this protocol, we require a simultaneous broadcast
channel. It would be particularly well-suited for a small group of
voters that are unwilling to trust any third party and who have no
advantage in making the protocol~abort.

Previous work on information-theoretically secure voting protocols
include~\cite{CFSY96}, where a protocol is given in the context
where many election authorities are present. To the best of our
knowledge, our approach is fundamentally different from any other
approaches for  voting. It is the first to provide
information-theoretic security without requiring or trusting any
third party, while also providing ballot casting assurance (each
participant is convinced that their input is correctly
recorded~\cite{AN06}) and universal verifiability (each participant
is conviced that only registered voters cast ballots and that the
tally is correctly computed~\cite{SK95}).

\subsubsection{Anonymity}

Anonymity is the power to perform a task without having to identify
the participants that are involved.  In the case of \emph{anonymous
message transmission}, it is simply the capacity of the sender to
transmit a private message to a specific receiver of his choosing
without revealing either his identity or the identity of the
receiver. A number of protocols have been suggested for anonymous
transmission. Many of these rely on trusted or semi-trusted third
parties as well as computational assumptions (for instance, the
MIX-net~\cite{Chaum81}). Here, we do not make any such assumptions.
The most notable protocol for anonymous transmission in our context
is the dining cryptographer's protocol~\cite{Chaum88}, which allows
a single sender to anonymously broadcast a bit, and provides
information-theoretical security against a passive adversary. We
present the protocol in a version that implements the multiparty
computation of the \emph{parity} function in
Section~\ref{sec:parity}.

The case of multiple yet honest senders in the dining
cryptographer's protocol can be solved by time slot reservation
techniques, as originally noted by Chaum~\cite{Chaum88}. But
nevertheless, any corrupt participant can jam the channel.
Techniques offering computational security to this problem have been
proposed~\cite{Chaum88,WP89}. Also,  computational assumptions allow
the removal of the reliance on a broadcast channel~\cite{WP89Disco}.

In our implementation of \emph{anonymous bit transmission}
(Section~\ref{sec:anonymous bit transmission}), we elegantly deal
with the case of multiple senders by allowing an unlimited amount of
participants to act as anonymous senders. Each anonymous sender can
target any number of participants and send them each a private bit
of his choice. Thus, the outcome of the protocol~is, for each
participant, a private list indicating how many~$0$s and how
many~$1$s were received. The anonymity of the sender and  receiver
and the privacy of all transmitted bits is always perfectly
achieved, but any participant can cause the protocol to abort, in
which case the participants may still learn some information about
their own private lists.

We need a way for all participants to find out if the protocol has
succeeded. This is done with the \emph{veto} protocol
(Section~\ref{sec:veto}), which takes as input a single bit from
each participant; the output of the protocol is the logical~OR of
the inputs. Our implementation differs from the ideal functionality
since a participant that inputs~$1$ will learn if some other
participant also input~$1$. We make use of this deviation from the
ideal functionality in further protocols.

In our \emph{fixed role anonymous message transmission} protocol
(Section~\ref{sec:fr-anonymous-message}), we present a method which
allows a single sender to communicate a message of arbitrary length
to a single receiver. To the best of our knowledge, this is the
first protocol ever to provide perfect anonymity, message privacy
and integrity. For a fixed security parameter, the anonymous message
transmission is asymptotically optimal.

Our final protocol  for~\emph{anonymous message transmission}
(Section~\ref{sec:anonymous-message}) allows a sender to send a
message of arbitrary length to a receiver of his choosing. While any
participant can cause the protocol to abort, the anonymity of the
sender and receiver is always perfectly achieved. The privacy of the
message is preserved except with exponentially small probability. As
far as we are aware, all previous proposed protocols for this task
require either computational assumptions or a majority of honest
participants. The protocol deals with the case of multiple senders
by first executing the \emph{collision detection} protocol
(Section~\ref{sec:collision}), in which each participant inputs a
single bit. The outcome only indicates  if the sum of the inputs
is~$0$, $1$ or more. Compared to similar protocols called \emph{time
slot reservation}~\cite{Chaum88,WP89}, our protocol does not leak
any additional information about the number of would-be senders. The
final protocol also makes use of the \emph{notification} protocol
(Section~\ref{sec:notification}) in which each participant chooses a
list of other participants that are to be notified. The output
privately reveals to each participant the logical~OR of his received
notifications. A special case of this protocol is when a single
participant notifies another single participant; this is the version
used in our final protocol to enable the sender to anonymously tell
to the receiver to act accordingly.

\subsection{Common features to all protocols}

All protocols presented in the following sections share some common
features, which we now describe. Our protocols are given in terms of
multiparty computation with inputs and outputs and involve~$n$
participants, indexed by~$i=1,\ldots ,n$. In the ideal
functionality, the only information that the participants learn is
their output (and what can be deduced from it). \emph{Correctness}
refers to the fact that the outputs are correctly computed, while
\emph{privacy} ensures that the inputs are never revealed.

The protocols ensure correctness and privacy even in the presence of
an unlimited number of misbehaving participants. Two types of such
behaviour are relevant: participants who collude (they follow the
protocol but pool their information in order to violate the
protocol's privacy), and participants who actively deviate from the
protocol (in order to violate the protocol's correctness or
privacy). Without loss of generality, these misbehaviours are
modelled by assuming a central adversary that controls  some
participants, rendering them \emph{corrupt}. The adversary is either
\emph{passive} (it learns all the information held by the corrupt
participants), or \emph{active} (it takes full control of the
corrupt participants). We will deal only with the most general case
of active adversaries, and require them to be~\emph{static} (the set
of corrupt participants does not change). A participant that is not
corrupt is called~\emph{honest}. Our protocols are such that if they
do not abort, there exists inputs for the corrupt participants that
would  lead to the same output if they were to act honestly. If a
protocol aborts, the participants do not learn any more information
than they could have learned in an honest execution of the protocol.
The input and output description applies  only to honest
participants.

We assume that each pair of participants shares a
private, uniformly random string that can be used to implement an
authentic private channel. The participants have access to a
broadcast channel and in some cases, it is simultaneous. A
\emph{broadcast} channel is an authentic broadcast channel for which
the sender is confident that all participants receive the same value
and the receivers know the identity of the sender. A
\emph{simultaneous broadcast} channel is a collection of broadcast
channels where the input of one participant cannot depend on the
input of any other participant. This could be achieved if all
participants {\em simultaneously} performed a broadcast. In order to
distinguish between the two types of broadcast, we sometimes call
the broadcast channel a \emph{regular} broadcast.
It is not
uncommon in multiparty computation to allow additional resources,
even if these resources cannot be implemented with the threshold on
the honest participants (the results of~\cite{RB89} which combine a
broadcast channel with~$n/2$ honest participants being the most
obvious example). Our work suggests that a simultaneous broadcast
channel is an interesting primitive to study in this context.

In all protocols, the security parameter is~$s$. Unfortunately, in
many of our protocols, a single corrupt participant can cause the
protocol to abort. All protocols run in polynomial time with respect
to the number of participants, the security parameter and the input
length. Although some of the protocols presented in this paper are
efficient, our main focus here is in the \emph{existence} of
protocols for the described tasks. We leave for future work
improvement of their efficiency. Finally, due to lack of space, we
present only sketches of security proofs.

\section{Parity}
\label{sec:parity}

Protocol~\ref{prot:parity} implements the \emph{parity} function
and is essentially the same as the dining cryptographers
protocol~\cite{Chaum88}, with the addition of a simultaneous
broadcast channel. Note that if we used a broadcast channel instead,
then the last participant to speak would have the unfair advantage
of being able to adapt his input in order to fix the outcome of the
protocol!

\begin{protocol} \caption{Parity} \label{prot:parity}
{\bf Input:} $x_i\in\{0,1\}$ \\
{\bf Output:} $y_i=x_1 \oplus x_2 \oplus \cdots  \oplus x_n$ \\
{\bf Broadcast type :} simultaneous broadcast \\
{\bf Achieved functionality:} \\
1) (Correctness) If the protocol does not abort, the output is the same as in the ideal functionality.  \\
2) (Privacy) No adversary can learn more than the output of the
ideal functionality. \vspace{2pt}
 \hrule  \vspace{2pt}
Each participant~$i$ does the
following:
\begin{enumerate}
\item \label{ParityStep1} Select uniformly at random an $n$-bit  string $r_i = r_i^1r_i^2\ldots r_i^n$ with Hamming weight of parity~$x_i$.
\item Send $r_i^j$ to participant $j$ using the private channel;
keep bit $r_i^i$ to yourself.
\item Compute $z_i$, the parity of the sum of all the bits  received, including~$r_i^i$.
\item Use the simultaneous broadcast channel to announce~$z_i$.
\label{step:broadcast}
\item \label{step:compute}After the simultaneous broadcast is finished, compute $y_i= \bigoplus_{k=1}^{n} z_k $. This is the outcome of the protocol.
If the simultaneous broadcast fails, abort the protocol.
\end{enumerate}
\end{protocol}

Correctness and privacy follows from~\cite{Chaum88}. Thus, any
adversary can learn only what can be deduced from the corrupt
participant's inputs and the outcome of the protocol. Note that this
means that the adversary can  deduce the parity of the inputs of the
other participants. We will later use the two simple observations
that there is no way to cheat except by refusing to broadcast and
that any value that is broadcast is consistent with a choice of
valid inputs. In the following protocols, we will adapt
step~\ref{step:broadcast} of the \textbf{parity} protocol  to make
it relevant to the scenario, this will allow us to remove the
assumption of the simultaneous broadcast. We will also use the fact
that if a single participant either does not broadcast, or
broadcasts a random bit in step~\ref{step:broadcast} then the value
of the output of \textbf{parity} is known to this participant, but
is perfectly hidden to all other participants.

\section{Veto}
\label{sec:veto}

In this section, we build on the \textbf{parity} protocol to give a
protocol for the secure implementation of the \emph{veto} function,
which computes the logical~OR of the participant's inputs
(Protocol~\ref{prot:veto}). We also discuss some important
deviations from the ideal functionality.
\begin{protocol} \caption{Veto} \label{prot:veto}
{\bf Input:} $x_i\in\{0,1\}$ \\
{\bf Output:} $y_i=x_1 \vee x_2 \vee \cdots  \vee x_n$ \\
{\bf Broadcast type :} regular broadcast\\
{\bf Achieved functionality:} \\
1) (Reliability) No participant can make the protocol abort. \\
2) (Correctness) The outcome of the protocol equals the outcome of the ideal functionality. \\
3) (Privacy) Any adversary learns what it would have learned in the
ideal functionally, had the corrupt participants used~$0$ as input,
but nothing more.

 \vspace{4pt} \hrule \vspace{4pt}

The~$n$ participants agree on~$n$ orderings such that each ordering
has a different last participant.

$\text{\textsf{result}} \leftarrow 0$

For each ordering,\\
\phantom{----}Repeat~$s$ times: \hspace{1cm}\begin{enumerate}
\item \label{step:flip}Each participant~$i$ sets the value of~$p_i$ in the
following way: if~$x_i=0$ then $p_i=0$; otherwise, $p_i=1$ with
probability~$\frac{1}{2}$ and $p_i=0$ with complimentary
probability.
\item \label{veto:step2}The participants execute the
\textbf{parity} protocol with inputs $p_1, p_2, \ldots p_n$, with
the exception that the simultaneous broadcast  is replaced by a
regular broadcast with the participants broadcasting according to
the current ordering (if any participant refuses to broadcast, set
the value $\textsf{result} \leftarrow 1$). If the outcome of
\textbf{parity} is~$1$, then set $\text{result} \leftarrow 1$\,.
\end{enumerate}
Output the value \textsf{result}.
\end{protocol}

\begin{lemma} \label{lemma:veto3}(Reliability)
No participant can make the \textbf{veto} protocol abort.
\end{lemma}

\begin{proof}
If a participant refuses to broadcast, it is assumed that the output
of the protocol is~$1$.
\end{proof}

\begin{lemma} \label{lemma:veto1}(Correctness)
If all participants in the \textbf{veto} protocol have input
$x_i=0$, then the protocol achieves the ideal functionality with
probability~1. If there exists a participant with input $x_i=1$ then
the protocol is correct with probability at least~$1-2^{-s}$.
\end{lemma}

\begin{proof}
The correctness follows by the properties of the \textbf{parity}
protocol, with the difference that we now have a broadcast channel
instead of a simultaneous broadcast channel. The case where all
inputs are~0 is trivial. Let~$x_i=1$ and suppose that the protocol
is executed until the ordering in which participant~$i$ speaks last.
Then with probability at least~$1-2^{-s}$, in step~\ref{veto:step2}
of \textbf{veto}, the output of the  protocol will be set to~$1$.
\end{proof}

\begin{lemma}\label{veto_HC}
\label{lem:veto4} (Privacy) In the \textbf{veto} protocol, the most
 an adversary can learn is the information
that it could have learned by assigning to all corrupt participants
the input~$0$. Additionally, this information is revealed, even to a
passive adversary,
 with
probability at least~$1-2^{-s}$.
\end{lemma}

\begin{proof}
This follows from the properties of the \textbf{parity} protocol:
for a given repetition, the adversary learns the parity of the
honest participants'~$p_i$'s, but nothing else.
 Because of the way that the $p_i$'s are chosen in
step~\ref{step:flip}, if for any repetition, this parity is odd, the
adversary concludes that at least one honest participant has
input~$1$,  and otherwise if all repetitions yield~$0$, then the
adversary concludes that with probability at least~$1-2^{-s}$, all
the honest participant's inputs are~$0$.  In all cases, this is the
only information that is revealed; clearly, it is revealed to any passive adversary, except with exponentially small probability.
\end{proof}

\section{Vote}
\label{sec:vote}

The participants now wish to conduct an $m$-candidate~\textbf{vote}.
The idea of Protocol~\ref{prot:vote} is simple. In the \textbf{veto}
protocol, each participant with input~$1$ completely randomizes his
input into the \textbf{parity} protocol, thus randomizing the output
of \textbf{parity}. By flipping the output of \textbf{parity}
 with probability only~$1/n$, the
probability of the outcome being odd becomes a function of the
number of such flips. Using repetition, this probability can be
approximated to obtain the exact number of flips with exponentially
small error probability. This can be used to compute the number of
votes for each candidate. Unfortunately, a corrupt participant can
randomize his bit with probability higher than~$1/n$, enabling him
to vote more than once. But  since a participant cannot derandomize
the parity, he cannot vote less than zero times. Verifying that the
sum of the  votes equals~$n$ ensures that all participants vote
exactly once. Note that the protocol we present is polynomial in~$m$
and not in the length of~$m$.

\begin{protocol}
 \caption{Vote} \label{prot:vote}

{\bf Input:} $x_i \in \{1,\ldots,m\}$ \\
{\bf Output:} for $k=1$ to $m$, $y[k]_i= \abs{\{x_j \mid x_j=k \}}$  \\
{\bf Broadcast type :} simultaneous broadcast\\
{\bf Achieved functionality:} \\
1) (Correctness) If the protocol does not abort, then there exists
an input~$x_i$ for
each corrupt participant such that the protocol achieves the ideal functionality. \\
2) (Privacy) Even if the protocol aborts, no adversary can  learn
more that what it would have learned by setting in the ideal
functionality~$x_i=1$ for all corrupt participants.

 \vspace{4pt} \hrule \vspace{4pt}

\textbf{Phase A}

For each candidate $k=1$ to~$m$,

\hspace{0.3cm} For $j=1$~to~$s$,
\begin{enumerate}
\item \label{step:flip-1}Each participant~$i$ sets the value of~$p_i$ in the
following way: if~$x_i \neq k$, then $p_i=0$; otherwise, $p_i=1$
with probability~$\frac{1}{n}$ and $p_i=0$ with complimentary
probability.
\item \label{step:parity} The participants execute the \textbf{parity} protocol to
compute the \emph{parity} of $p_1, p_2, \ldots p_n$, but instead of
broadcasting their output bit~$z_i$, they store it as~$z[k]_i^j$.
\end{enumerate}

\textbf{Phase B}

All participants simultaneously broadcast~$z[k]_i^j$~$(j=1,2,\ldots,
s)$.
If the simultaneous broadcast is not successful, the protocol aborts. \\

\vspace{-.4cm}

\textbf{Phase C}

 To compute the tally, $y[k]_i$, for each value
~$k=1 \ldots m$, each participant sets:
\mbox{$p[k]_j=\bigoplus_{i=1}^n z[k]_i^j$}, \mbox{$\sigma[k]_i =
\sum_{j=1}^s p[k]_j/s$} and  if there exists an integer~$v$ such
that $\left|\sigma[k]_i - p_v \right| < \frac{1}{2e^2n} \,$, \\
where $p_v = \frac{1}{2}\left(\frac{n-2}{n}\right)^v
\left(\left(\frac{n}{n-2}\right)^v -1\right)$, then $y[k]_i=v$\,.

If for any~$m$, no such value~$v$ exists, or if $\sum_{k=1}^m y[k]_i
\neq n$, the protocol aborts.

\end{protocol}

\begin{lemma}(Correctness)
\label{lem:correctness corrupt} If the \textbf{vote} does not abort,
then there exists an input for each corrupt participant such that
the output of the honest participants equals the output of the ideal
functionality, except with probability exponentially small in~$s$.
\end{lemma}

\begin{proof}
If all participants are honest, the correctness of the protocol is
derived from the Chernoff bound  as explained in the Appendix.
Assume now~$t$ corrupt participants. Since the \textbf{parity}
protocol is perfect, the only place participant~$i$ can deviate from
the protocol is by choosing~$p_i$ with an inappropriate probability.
We first note that if the~$t$ corrupt participants actually transmit
the correct number of private bits in \textbf{phase~A} and broadcast
the correct number of bits in \textbf{phase~B}, then whatever they
actually send is consistent with some global probability of
flipping.

We use again the fact that it is possible to randomize the parity
but not to derandomize it: if the corrupt participants altogether
flip with a probability not consistent with an integer number of
votes, either the statistics will be inconsistent, causing the
protocol to abort, or we can interpret the results as being
consistent with an integer amount of votes. If they flip with a
probability consistent with an integer different than~$t$, then
each~$y[k]_i$ will  be assigned a value, but with probability
exponentially close to~1, we will have~$\sum_{k=1}^m y[k]_i \neq n$
and the protocol will~abort.
\end{proof}

\begin{lemma}(Privacy)
\label{lem:privacy} In the \textbf{vote} protocol, no adversary can
learn more than what it would have learned by assigning to all
corrupt participants the input~$1$ in the ideal functionality, and
this even if the protocol aborts.
\end{lemma}

\begin{proof}
Assume that the first~$t$ participants are corrupt. No information
is sent \textbf{phase~A} or \textbf{phase~C}. We thus have to
concentrate on \textbf{phase~B} where the participants broadcast
their information regarding each parity. For each execution of
\textbf{parity}, the  adversary learns the parity of the honest
participant's values, $p_{t+1} \oplus p_{t+2} \oplus \ldots \oplus
p_{n}$, but no information on these individual values is revealed.
The adversary can thus only evaluate the probability with which the
other participants have flipped the parity. But this information
could be deduced from the output of the ideal functionality, for
instance by fixing the corrupt participants' inputs to~$1$.
\end{proof}

It is important to note that the above results do not exclude the
possibility of an  adversary causing the protocol to abort while
still learning some information as stipulated in
Lemma~\ref{lem:privacy}. This information could be used to adapt the
behaviour of the adversary in a future execution of~\textbf{vote}.

In addition to the above theorems, it follows from the use of the
simultaneous broadcast channel that an adversary cannot act in a way
that a corrupt participant's vote depends an honest participant's
vote. In particular, it cannot \emph{duplicate} an honest
participant's vote. We claim that our protocol provides ballot
casting assurance and universal verifiability. This is
straightforward from the fact that participants do not entrust any
computation to a third party: they provide their own inputs and can
verify that the final outcome is computed correctly.

\section{Anonymous Bit Transmission}
\label{sec:anonymous bit transmission}

The \textbf{anonymous bit transmission} protocol enables a sender to
privately and anonymously transmit one bit to a receiver of his
choice. Protocol~\ref{prot:anon-bit-trans} actually deals with the
usually problematic scenario of multiple \emph{anonymous senders} in
an original way: it allows an arbitrary number participants to act
as anonymous senders, each one targeting any number of participants
and sending them each a chosen private bit.  Each participant is
also simultaneously a potential \emph{receiver}: at the end of the
protocol, each participant has a private account of how many
anonymous senders sent the bit~$0$ and how many sent the bit~$1$.
Note that in our formalism for multiparty computation, the
\emph{privacy }of the inputs implies the \emph{anonymity} of the
senders and receivers.

\begin{protocol}
\caption{Anonymous Bit Transmission} \label{prot:anon-bit-trans}
{\bf Input:} $ x^j_i \in \{0,1,\perp \}$, $(j=1,2,\ldots,n)$ \\
{\bf Output:} $y_i= ( \abs{\{x_j^i \mid x_j^i=0 \}} ,   \abs{\{x_j^i \mid x_j^i=1\} })$ \\
{\bf Broadcast type :}  regular broadcast\\
1) (Correctness) If the protocol does not abort then the output of
the protocol
equals the output of the ideal functionality.   \\
2) (Privacy) The privacy is the same as in the ideal functionality.

 \vspace{4pt} \hrule \vspace{4pt}

For each participant~$j$,
\begin{enumerate}
\item Execute the \textbf{vote} protocol with~$m=3$ as modified below. The three choices are: $0$, $1$, or
$\perp$ (\emph{abstain}). Each participant~$i$ chooses his input to
the \textbf{vote} according to $x^j_i$, his choice of message to be
sent anonymously to participant~$j$. The \textbf{vote} protocol is
modified such that:
\begin{enumerate}
\item \label{Anon:broadcast}The output strings are sent to participant~$j$
through the private channel.\item Participant~$j$ computes the tally
as in the \textbf{vote}  and if this computation succeeds, he finds
out how many participants sent him a~$0$, how many sent him a~$1$
and how many abstained. If this occurs (and the results are
consistent) he sets his success bit,~$s_j$ to~$0$. If the
\textbf{vote} aborts, he sets~$s_j$ to~$1$.
\end{enumerate}
\end{enumerate}
Execute the \textbf{veto} protocol, using as inputs the success
bits~$s_j$. If the output of \textbf{veto} is~$0$, then the
\textbf{anonymous bit transmission} succeeds. Otherwise, the
protocol fails.
\end{protocol}

The security of the \textbf{anonymous bit transmission} protocol
follows directly from the security of the \textbf{vote}  and
of the \textbf{veto}. Of course, the \textbf{anonymous bit
transmission} also inherits the drawbacks of these protocols. More
precisely we have the following:

\begin{lemma}(Correctness)
The \textbf{anonymous bit transmission} protocol computes the
correct output, except with exponentially small probability.
\end{lemma}

\begin{proof}
If the protocol does not abort, by Lemmas~\ref{lemma:veto1}
and~\ref{lem:correctness corrupt}, except with exponentially small
probability, all bits are correctly transmitted.
\end{proof}

\begin{lemma}(Privacy)
In the \textbf{anonymous bit transmission} protocol, the privacy is
the same as in the ideal functionality.
\end{lemma}

\begin{proof}
Each execution of the \textbf{vote} protocol  provides perfect
privacy, even if the protocol aborts.
 The final veto reveals some partial information about which honest participants
have been targeted by corrupt participants, but this does not
compromise the privacy of the protocol.\end{proof}

In Protocol~\ref{prot:anon-bit-trans}, the use of the private
channel in step~(a) can be removed and replaced by a broadcast
channel. Since participant~$j$ does not broadcast, the messages
remain private. Another  modification of the protocol makes it
possible to send~$m$ possible messages instead of just two but note
that the complexity is polynomial in~$m$ and not in the length
of~$m$. The transmission of arbitrarily long strings is discussed in
Sections~\ref{sec:fr-anonymous-message}
and~\ref{sec:anonymous-message}.

\section{Collision Detection}
\label{sec:collision}

The \textbf{collision detection} protocol
(Protocol~\ref{prot:collision-detection}) enables the participants
to verify whether or not there is a single sender in the group. This
will be used as a procedure for the implementation of
\emph{anonymous message transmission} in
Section~\ref{sec:anonymous-message}. Ideally, a protocol to detect a
collision would have as input only~$x_i \in \{0,1\}$ but
unfortunately we do not know how to achieve such a functionality.

\begin{protocol}
\caption{Collision Detection} \label{prot:collision-detection}

{\bf Input:} $x_i \in \{0,1,2\}$ \\
{\bf Output:} let $r=\sum_{i=1}^n x_i$ then~$y_i=\min\{r,2\}$  \\
{\bf Broadcast type :}  regular broadcast\\
1) (Reliability) No participant can make the protocol abort.\\
2) (Correctness) The output of the protocol equals the output of
the ideal functionality.\\
3) (Privacy)  An adversary cannot learn more than it could have
learned by assigning to all corrupt participants the input~$0$ in
the ideal functionality.

\vspace{4pt} \hrule \vspace{4pt}

{\bf Veto~A}

All participants perform the \textbf{veto} protocol with inputs
$\min\{x_i,1\}$.\\

\vspace{-.4cm}

{\bf Veto~B}

If the outcome of \textbf{veto~A} is~$0$, skip this step. Otherwise,
each participant with input~1 in \textbf{veto~A} will set~$b_i=1$ if
he detected in \textbf{veto~A} that another participant had
input~$1$, or if~$x_i=2$. All other participants set~$b_i=0$. Then
all participants perform a second \textbf{veto} protocol with
inputs~$b_i$.\\

\vspace{-.4cm}

{\bf Output:} $\ y_i = \begin{cases} 0 &\text{if the
outcome of \textbf{veto~A} is~0}\\
1 &\text{if the outcome of \textbf{veto~A} is~1 and the outcome of
\textbf{veto~B}
is~0 }\\
2  &\text{if the outcome of \textbf{veto~A} is~1 and the outcome of
\textbf{veto~B} is~1}
\end{cases}$

\end{protocol}

\begin{lemma}(Reliability)
No participant can make the \textbf{collision detection} protocol
abort.
\end{lemma}
\begin{proof}
This follows from the reliability of \textbf{veto}.
\end{proof}

\begin{lemma}(Correctness)
In the \textbf{collision detection} protocol, the output equals the
output of the ideal functionality (except with exponentially small
probability).
\end{lemma}

\begin{proof}
This follows from the correctness of the \textbf{veto} protocol.
There are only two ways a corrupt participant can deviate from the
protocol. First, participant~$i$ can set~$b_i = 0$ although~$x_i \in
\{0,1\}$ and although in the first veto his input was~1 and a
collision was detected. The outcome of \textbf{veto~B} will still
be~1 since another participant with input~1 in \textbf{veto~A} will
input~1 in \textbf{veto~B}. This is consistent with input~$x_i=1$.
Second, participant~$i$ can set~$b_i =1$ although~$x_i=0$. If
\textbf{veto~B} is executed, then we know that another participant
has input~1 in \textbf{veto~A}.
 This is  consistent with input $x_i=1$. \end{proof}

Note that we have raised a subtle deviation from the ideal protocol
in the above proof:   we showed how it is possible for a corrupt
participant to set his input to~0 if all other participants have
input~$0$ and to~$1$ otherwise. Fortunately, the protocol is still
sufficiently good for the requirements of the following sections.

\begin{lemma}(Privacy)
In the \textbf{collision detection}  protocol, an adversary cannot
learn more than it could have learned by assigning to all corrupt
participants the input~$0$ in the ideal functionality.
\end{lemma}

\begin{proof}
In each \textbf{veto}, an adversary  can only learn whether or not
there exists an honest participant with input~1. In all cases, this
can be deduced from the outcome of the ideal functionality by
setting the input to be~$0$ for all corrupt participants.
\end{proof}

\section{Notification}
\label{sec:notification}

In the \textbf{notification} protocol
(Protocol~\ref{prot:notification}), each participant chooses a list
of other participants to notify. The output privately reveals to
each participant whether or not he was notified, but no information
on the number or origin of such notifications is revealed.

\begin{protocol} \caption{Notification} \label{prot:notification}
{\bf Input:} $\forall j \neq i, x^j_i \in \{0,1\}$ \\
{\bf Output:} $y_i=\bigvee_{j \neq i}  \ x_j^i$ \\
{\bf Broadcast type :}  regular broadcast\\
1) (Correctness) If the protocol does not abort then the output of
the protocol
equals the output of the ideal functionality.   \\
2) (Privacy) The privacy is the same as in the ideal functionality.

\vspace{4pt} \hrule \vspace{4pt}

For each participant~$i$:\\
\phantom{----}Participant~$i$ sets $y_i \leftarrow 0$.\\
\phantom{----}Repeat~$s$ times:
\begin{enumerate}
\item Each participant~$j \neq i$ sets the value of~$p_j$ in the
following way: if~$x_j^i=0$ then $p_j=0$; otherwise, $p_j=1$ with
probability~$\frac{1}{2}$ and $p_i=0$ with complimentary
probability. Let~$p_i =0$.
\item The participants execute the \textbf{parity} protocol with inputs
$p_1, p_2, \ldots p_n$, with the exception that participant~$i$ does
not broadcast his value, and the simultaneous broadcast is replaced
by a regular broadcast (if any participant refuses to broadcast,
abort).
\item Participant~$i$ computes the outcome of \textbf{parity}, and if it is~$1$, $y_i \leftarrow
1$\,.
\end{enumerate}
\end{protocol}

\begin{lemma}
The \textbf{notification} protocol achieves privacy and except with
exponentially small probability, the correct output is computed.
\end{lemma}

\begin{proof}
Privacy and correctness are trivially deduced from properties of the
\textbf{parity} protocol.
\end{proof}

\section{Fixed Role Anonymous Message Transmission}
\label{sec:fr-anonymous-message}

In Section~\ref{sec:anonymous bit transmission}, we presented an
\textbf{anonymous bit transmission} protocol. The protocol easily
generalizes to~$m$ messages, but the complexity of the protocol
becomes polynomial in~$m$. It is not clear how to modify the
protocol to transmit a string of arbitrary length, while still
allowing multiple senders and receivers. However, in the context
where a single sender~$S$ is allowed, it is possible to implement a
secure protocol for~$S$ to anonymously transmit a message to a
single receiver~$R$, which we call \textbf{fixed role anonymous
message transmission} (Protocol~\ref{prot:fr-anon-mes-trans}). If
the uniqueness condition on~$S$ and~$R$ is not satisfied, the
protocol aborts.  The protocol combines the use of the
\textbf{parity} protocol with an \emph{algebraic manipulation
detection code}~\cite{CFP07}, which we present as
Theorem~\ref{thm:serge}. Due to lack of space, the encoding and
decoding algorithms,  $F$ and~$G$, respectfully, are not repeated.
For a less efficient algorithm that achieves a similar result,
see~\cite{PSV99}.

\begin{theorem}[\cite{CFP07}] \label{thm:serge}
There exists an efficient probabilistic encoding algorithm
$F:\{0,1\}^m \rightarrow \{0,1\}^{m+2(\log(m)+s}$ and decoding
algorithm $G:\{0,1\}^{m+2(\log(m)+s)} \rightarrow
\{\perp,\{0,1\}^m\}$ such that for all~$w$, $G(F(w))=w$, and
any combination of bit
flips applied to~$w'=F(w)$ produces a~$w''$ such that
$G(w'')=\perp$, except with probability~$2^{-s}$.
\end{theorem}

\begin{protocol}\caption{Fixed Role Anonymous Message Transmission}
\label{prot:fr-anon-mes-trans}
{\bf Oracle:} The sender~$S$ and receiver~$R$ know their identity \\
{\bf Input:} $S$ has input~$w \in \{0,1\}^m$, all other players have no input  \\
{\bf Output:} $R$ has output~$w$, all other players have no output    \\
{\bf Broadcast type :}  regular broadcast\\
1) (Correctness) If the protocol does not abort, $R$ obtains the
correct message. \\
2) (Privacy) The only information that can be learned through the
protocol
is for~$R$ to learn~$w$. \\
3) (Oracle) If the oracle conditions are not satisfied (in the sense
that more than one honest participant  believes to be the sender or
the receiver), the protocol will abort.

\vspace{4pt} \hrule \vspace{4pt}

\begin{enumerate}
\item $S$ computes~$w'=F(w)$
\item The participants execute~$m+2(\log (m)+s)$ rounds of the  \textbf{parity}
protocol, with participants using a
broadcast instead of a simultaneous broadcast and using the
following inputs:
\begin{enumerate}
\item $S$ uses as input the bits of~$w'$.
\item $R$ uses as input the bits of a random $m$-bit string,~$r$.
\item All other players use $0$ as input for each round.
\end{enumerate}
\item Let $d$ be the output of the rounds of \textbf{parity}.  $R$~computes~$w''=d \oplus r$.
\item $R$ computes $y=G(w'')$.
\item \label{step:veto} A \textbf{veto} is performed:  all players input~0 except $R$ who inputs~$1$ if  $y=\perp$ and~$0$ otherwise.  \\
If the outcome of \textbf{veto} is~1, the protocol aborts.
Otherwise, $R$ sets his output to~$y$.
\end{enumerate}
\end{protocol}

\begin{lemma}(Correctness, Privacy, Oracle)
In the \textbf{fixed role anonymous message transmission} protocol,
the probability that~$R$ obtains as output a corrupt message is
exponentially small. The protocol is perfectly private,  and  if the
oracle conditions are not satisfied, it will abort (except with
exponentially small probability).
\end{lemma}

\begin{proof}
Because of the properties of \textbf{parity} and the fact that the
receiver broadcasts a random bit, we have perfect privacy.
Correctness is a direct consequence of Theorem~\ref{thm:serge}.
Finally, if more than one participant acts as a sender or
receiver, then again because of Theorem~\ref{thm:serge}, the message
will not be faithfully transmitted and the protocol will abort in
step \ref{step:veto}, except with  exponentially small probability.
\end{proof}

\begin{theorem} \label{thm:optimal}
For a fixed security parameter, the \textbf{fixed role anonymous
message transmission} protocol is asymptotically optimal.
\end{theorem}

\begin{proof}
For any protocol to preserve the anonymity of the sender and the
receiver,  each player must sent at least one bit to every other
player for each bit of the message. In the  \textbf{fixed role
anonymous message transmission} protocol, for a fixed~$s$,  each
player actually sends $O(1)$ bits to each other player and therefore
the protocol is asymptotically optimal.
\end{proof}

\section{Anonymous Message Transmission}
\label{sec:anonymous-message}

Our final protocol allows a sender to anonymously transmit
message to a receiver of his choosing. Contrary to  the
\textbf{fixed role anonymous message transmission} protocol of
Section~\ref{sec:fr-anonymous-message}, \textbf{anonymous message
transmission} (Protocol~\ref{prot:anon-mess-trans}) does not suppose
that there is a single sender, but instead, it deals with potential
collisions (or lack of any sender at all) by producing the outputs
\textsc{Collision} or \textsc{No~Transmission}. The only deviation
from the ideal functionality in the  protocol is that a single
participant can force the \textsc{Collision} output. Note again that
in this protocol, the privacy of the input implies anonymity of the
sender and receiver.

\begin{protocol}
\caption{Anonymous Message Transmission}
\label{prot:anon-mess-trans}
{\bf Input:} $x_i=\perp$ or $x_i = (r,w)$ where $r \in \{1,\ldots,n\}$ and  $w \in \{0,1\}^m$ \\
{\bf Output:} If $|\{x_i \mid x_i \neq  \perp \}|=0$ then
$y_i=\textsc{No~Transmission}$ and if $|\{x_i \mid x_i \neq  \perp
\}|>1$
 then $y_i=\textsc{Collision}$. Otherwise let~$S$ be such that $x_S=(r,w)$ then all $y_i=\perp$ except $y_r=w$. \\
{\bf Broadcast type :}  regular broadcast \\
1) (Correctness) The output equals the output of the ideal
functionality except that a single participant can make the protocol produce the output \textsc{Collision}.\\
2) (Privacy) The privacy is the same as in the ideal functionality.

\vspace{4pt} \hrule \vspace{4pt}

\begin{enumerate}
\item The participants execute the \textbf{collision detection} protocol; participants who have input~\mbox{$x_i
=\perp$} use input~$0$ while all others use input~$1$. If the
outcome of \textbf{collision detection} is~$1$, continue, otherwise
output \textsc{No~Transmission} if the output is~$0$ and
\textsc{Collision} if the output is~$2$.

\item \label{anon-mess-trans-step-2} Let the sender~$S$ be the unique participant with~$x_S \neq
\perp$. The participants execute the \textbf{notification} protocol,
with~$S$ using input~$x_S^r =1$ and $x_S^j=0$ otherwise. All other
participants use the input bits~$0$. Let~$R$ be the participant who
computes as output~$y_R = 1$. If the \textbf{notification} protocol
fails, abort.

\item The participants execute the \textbf{fixed role anonymous message transmission}
protocol.

\end{enumerate}

\end{protocol}

\begin{lemma}(Correctness)
In the \textbf{anonymous message transmission} protocol, the output
equals the output of the ideal functionality except with
exponentially small probability. The only exception is  that a
single participant can make the protocol produce the output
\textsc{Collision}.
\end{lemma}

\begin{proof}
This follows easily from the correctness of the \textbf{collision
detection}, \textbf{notification} and \textbf{fixed role anonymous
message transmission} protocols.
\end{proof}

\begin{lemma}(Privacy)
The anonymity of the sender and receiver are perfect. If the
protocol succeeds, except with exponentially small probability,
participant~$r$ is the only participant who knows~$w$.
\end{lemma}
\begin{proof}
Perfect anonymity follows from the privacy of the \textbf{collision
detection}, \textbf{notification} and \textbf{anonymous message
transmission} protocols. If the sender successfully notifies the
receiver in step~\ref{anon-mess-trans-step-2}, then the privacy
of~$w$ is perfect. But with exponentially small probability, the
receiver will not be correctly notified, and an adversary acting as
the receiver will successfully receive the message~$w$.
\end{proof}

\section{Conclusion}

We have given six multiparty protocols that are
information-theoretically secure without any assumption on the
number of honest participants. It would be interesting to see if the
techniques we used can be applied to other multiparty functions or
in other contexts.

Our main goal was to prove the existence of several protocols in a
model that does not make use of any strong hypotheses such as
computational assumptions or an honest majority. This being said,
all the presented protocols are reasonably efficient: they are all
polynomial in terms of communication and computational complexity
and in one case, asymptotically optimal.

\section*{Acknowledgements}
The authors wish to thank Hugue Blier, Gilles Brassard, Serge Fehr
and S\'ebastien Gambs. A.\,B.\ is supported by scholarships from the
\textsc{CFUW}, \textsc{FQRNT} and  \textsc{NSERC}. A.\,T.\ is
supported by \textsc{CIFAR}, \textsc{FQRNT}, \textsc{MITACS} and
\textsc{NSERC}.

\newpage

\bibliographystyle{alpha}
\bibliography{BT07a}

\appendix{}
\section{Proof of correctness for Protocol~\ref{prot:vote}}

\begin{lemma}(Correctness)
 If all participants are honest in  the \textbf{vote} (Protocol~\ref{prot:vote}),
then the output is correct, except with probability exponentially
small in~$s$.
\end{lemma}

\begin{proof}
We fix a value~$k$ and suppose that~$v$ participants have
input~$x_i=k$. Thus we need to show that  in the \textbf{vote},
$y[k]_i = v$,  except with probability exponentially small in~$s$.

We now give the  intuition behind \textbf{phase~C} of the
\textbf{vote}. Let~$p_v$ be the probability that
$p[k]_j=\bigoplus_{i=1}^n z[k]_i^j = 1$. For  $v \leq n$, we
have~$p_0 = 0$, $p_1 = \frac{1}{n}$ and $p_{v+1} =
p_v\left(1-\frac{1}{n}\right) + (1-p_v)\frac{1}{n}$. Solving this
recurrence, we get
\begin{equation} p_v =
\frac{1}{2}\left(\frac{n-2}{n}\right)^v
\left(\left(\frac{n}{n-2}\right)^v -1\right) \,. \end{equation}

Thus, the idea
 of \textbf{phase~C} of the \textbf{vote} is for the participants to approximate~$p_v$ by computing
\mbox{$\sigma[k]_i = \sum_{i=1}^s p[k]_j/s$}. If the approximation
is within~$\frac{1}{2e^2n}$ of~$p_v$, then the outcome
is~$y[k]_i=v$. We first show that if such a~$v$ exists, it is
unique.

Clearly,  for $v< n$, we have that $p_{v+1} > p_v$. We also have
$\lim_{n \rightarrow \infty}p_n = \frac{1}{2} - \frac{1}{2e^2}$.
Thus the difference between $p_{v+1}$ and $p_v$ is:
\begin{align}
p_{v+1} - p_v &= p_v\left(1-\frac{1}{n}\right) + (1-p)\frac{1}{n}
-p_v \\
&= \frac{1-2p_v}{n} \\
&> \frac{1-2p_n}{n} \\&>  \frac{1}{e^2n} \end{align}

Hence if
 such a~$v$ exists, it is unique. We now show
that except with probability exponentially small in~$s$, the
correct~$v$ will be chosen. Let $X= \sum_{j=1}^s p[k]_j$ be the sum
of the~$s$ executions of \textbf{parity}, with~$\mu=sp_v$ the
expected value of~$X$. The participants have computed  $\sigma[k]_i
= X/s$\,.

 By the
Chernoff bound, for any~\mbox{$0 < \delta \leq 1$},
\begin{equation}
\Pr[X \leq (1- \delta)\mu] < \exp(-\mu \delta^2/2)
\end{equation}
Let $\delta = \frac{1}{2e^2np_v}$. We have
\begin{equation}
\Pr[X \leq \mu - \frac{s}{2e^2n}]  < \exp(-\frac{s}{8e^4n^2p_v})
\end{equation}
and so
\begin{equation}
\Pr[\sigma[k]_i-p_v \leq \frac{-1}{2e^2n}] <
\exp(-\frac{s}{8e^4n^2p_v})
\end{equation}

 Similarly, still by the Chernoff
bound, for any~\mbox{$\delta < 2e -1$},
\begin{equation}
\Pr[X > (1+\delta) \mu] < \exp(-\mu \delta^2/4)
\end{equation}
Let $\delta = \frac{1}{2e^2np_v}$ and we get
\begin{equation}
\Pr[X >  \mu + \frac{s}{2e^2n}] < \exp(\frac{-s}{16e^4n^2p_v})
\end{equation}
and so
\begin{equation}
\Pr[\sigma[k]_i-p_v > \frac{1}{2e^2n}] <
\exp(\frac{-s}{16e^4n^2p_v})
\end{equation}

Hence the protocol produces the correct value for~$y[k]_i$, except
with probability exponentially small in~$s$.
\end{proof}

\end{document}